# Achievement of Ultra-High Quality Factor in Prototype Cryomodule for LCLS-II

G. Wu[1], A. Grassellino, E. Harms, N. Solyak, A. Romanenko, C. Ginsburg, R. Stanek

*Fermi National Accelerator Laboratory, Batavia, Illinois, 60510, USA*

Quality factor is a primary cost driver for high energy, continuous wave (CW) SRF linacs like the LCLS-II X-ray free electron laser currently under construction. Taking this into account, several innovations were introduced in the LCLS-II cryomodule design to push substantially beyond the previous state-of-the-art quality factor achieved in operation. This includes the first ever implementation of the nitrogen doping cavity treatment, the capability to provide high mass flow cooldown to improve expulsion of magnetic flux based on recent R&D, high performance magnetic shielding, and other critical subcomponents. To evaluate the implementation of these new cryomodule features, two prototype cryomodules were produced. In this paper, we present results from the prototype cryomodule assembled at Fermilab, which achieved unprecedented cavity quality factors of $3.0 \cdot 10^{10}$ at a nominal cryomodule voltage. We overview cavity performance, procedures to achieve ambient magnetic field < 5 mG at the cavity wall, and the successful demonstration of high mass flow cooldown in a cryomodule. The cavity performance under various cool down conditions are presented as well to show the impact of flux expulsion on $Q_0$.

## I. Introduction:

Free electron laser (FEL) light sources such as LCLS and the European XFEL provide unique, unprecedented capabilities in a range of research areas[1, 2]. LCLS-II at SLAC will be the first CW X-ray FEL, providing high brightness with high repetition rate[3] through the use of superconducting RF technology. Operating an SRF linac as powerful as LCLS-II in CW mode results in significant heat dissipation to cryogenic temperatures by RF currents in the walls of the superconductor. As a result, cryogenic infrastructure and cryogenic operations can represent a relatively high fraction of the overall cost of the accelerator, and improvements in the quality factor ($Q_0$) of the cavities have a dramatic impact on the overall project cost. Therefore, a number of innovative ideas were implemented for the first time in LCLS-II, including in the cavity treatment and cryomodule design.

LCLS-II is the first large-scale application of nitrogen doping. Nitrogen doping in niobium cavities has been shown to increase quality factors by up to a factor of four compared to non-doped cavities[4]. This was first demonstrated in vertical test, and then implemented in a cryomodule-like environment to show readiness for applications[5]. Once LCLS-II began to prepare for production, the technology was transferred to the industry[6]. Large scale production is now in progress in the industry to produce cavities for LCLS-II. Prior to the completion of the first industrial-produced cavities, Fermilab produced nitrogen doped 9-cell cavities that were used to build two LCLS-II pre-production cryomodules, one at Fermilab and another one at Jefferson Lab.

In addition to nitrogen doping, the LCLS-II cryomodule design incorporated extremely robust magnetic shielding to keep the ambient magnetic field as low as possible to minimize degradation of the quality factor due to trapped flux[7]. In addition, the cryomodules were specially designed to improve the heat load capacity of cavity helium vessel and to allow for fast cool down of niobium cavities to reduce the magnetic field trapping even further[8].

The first prototype cryomodule incorporated the CW design changes and was assembled at Fermilab using eight nitrogen doped 9-cell cavities. The cryomodule was tested at 2 K and showed record $Q_0$, demonstrating the success of the changes to cavity treatment and cryomodule design.

## II. Cryomodule Instrumentation

Each LCLS-II cryomodule consists of eight cavities. Each cavity has two High Order Mode (HOM) couplers which uses RF feedthrough to extract HOM power to external load. Two layers of cold magnetic shield enclose most of the cavity that resulted in unprecedentedly low remnant field at cavity wall[7]. For the prototype cryomodule, four cavities were selected to be equipped with fluxgate magnetometers and Cernox™ type temperature sensors as illustrated in FIG. 1 and FIG. 2. Each of those four cavities has four temperature sensors attached on top and bottom of cell #1 and cell #9. One magnetometer sensor was mounted at the bottom of cell #1 to measure transverse field normal to the illustration. Another magnetometer sensor was mounted on top of the cavity cell #1, tilted 45-degree off the beam axis as shown in FIG. 2. This sensor is expected to detect the combination of horizontal and vertical components of the remnant magnetic field. Both sensors are single-axis magnetometers.

Additional temperature sensors monitor the beam pipe temperature, HOM coupler body, HOM coupler feedthrough and inner layer of magnetic shield. Five fluxgate magnetometers were also installed between the two layers of magnetic shield at cavity position one, four, five, six and eight. Before the cool down of the cryomodule, it was demagnetized[9]. After demagnetization, all eight internal

---

[1] Electronic mail: genfa@fnal.gov

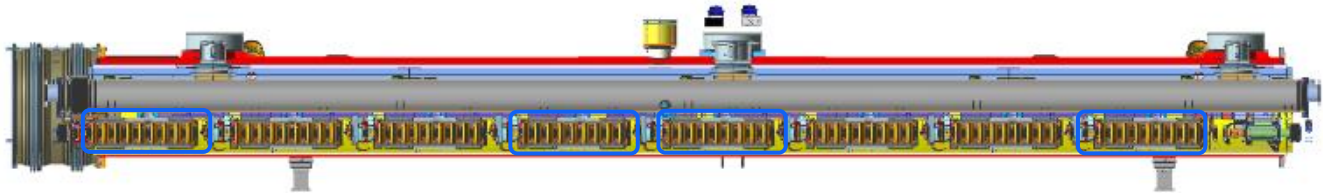

FIG. 1. A cross section view of an LCLS-II cryomodule with four cavities highlighted that have built-in magnetometers and temperature sensors.

magnetometers and five external magnetometers read less than one milligauss.

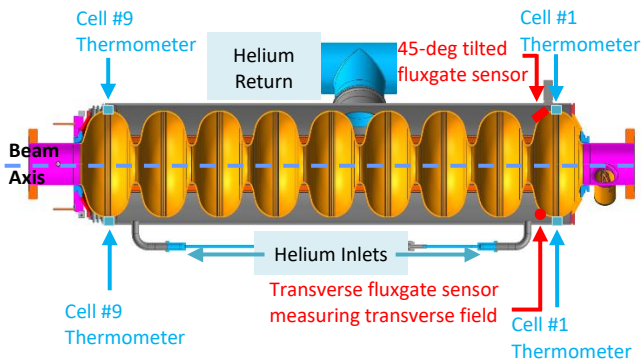

FIG. 2. A cross section view of a cavity that has built-in magnetometers and temperature sensors.

## III. Cryomodule Cool Down

The cryomodules was cooled at a rate of ~10 K per hour until cavities reached 45 K. The rate was increased to ~20 K per hour once the cavities crossed 100 K to minimize the risk of Q-disease (though cavities were hydrogen degassed during the processing which is expected to mitigate this degradation). Around 45 K, the cryomodule temperature was maintained through a soaking period of 24 hours to allow various components to continue their cool down including the magnetic shield. FIG 3 illustrates the one cavity temperature profile during cool down together with the temperatures of helium supply line and gaseous helium return pipe.

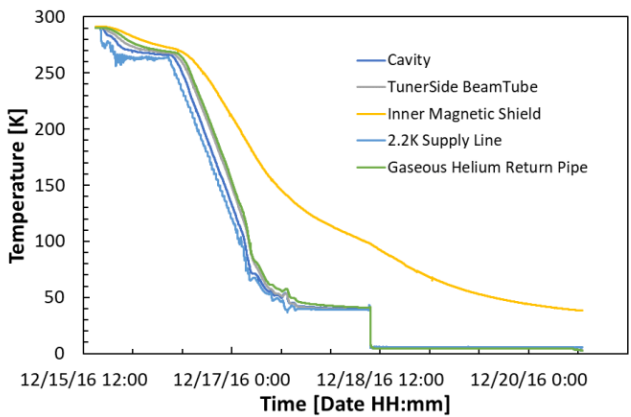

FIG. 3. Temperature profile during the cryomodule cool down.

**Fast Cool Down**

Once temperatures had settled, as shown in FIG 3, a fast cool down is initiated. A mass flow of approximately 80 g/s was recorded during the fast cool down. FIG. 4 illustrates the cavity temperature and fluxgate magnetometer readings during the fast cool down. For cavity one, cell #1 went through transition earlier than cell #9. The average temperature difference from top to bottom was 4.1 K.

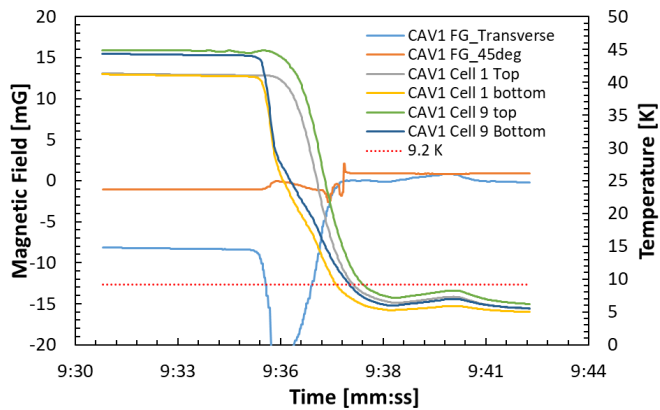

FIG. 4. Temperature and magnetic field profile during the fast cool down.

Table I lists the temperature difference for all four cavities. Despite the presence of a 50 K soaking period, magnetometers read the transverse remnant field ranged from 2 mG to 8 mG as compared to less than 1 mG readings prior to cool down. This indicates that the cryogenic circuit in the cryomodule has thermo-electric current due to differing Seebeck coefficients in various metals comprising closed electrical circuits[10], such as titanium-niobium or titanium-stainless steel.

Based on horizontal tests[7,11] and flux studies[12], the temperature gradient from the top of the cavity to the bottom is expected to be sufficiently large for reasonably strong flux expulsion per Table I.

TABLE I. Temperature differences and magnetic fields for four cavities during fast cool down. $B_T$ is the field measured by the transverse sensor and $B_{45}$ is the field measured by the 45-degree sensor.

| Cavity number | Temperature Difference [K] | $B_T$ [mG] | $B_{45}$ [mG] |
|---|---|---|---|
| 1 | 4.1 | 0.02 | 0.9 |
| 4 | 4.3 | 6.8 | 1.2 |
| 5 | 6.8 | 7.9 | 0.4 |
| 8 | 5.1 | 1.2 | 0.7 |

### A. Slow Cool Down

A slow cool down was implemented to evaluate the remnant field and its effect on the cavity quality factor. Cavity temperatures were raised to ~25 K before they were cooled to below the superconducting transition temperature. The cavity temperature showed very uniform temperature as indicated in FIG 5. The temperature difference was less than 0.08 K among the cells while the end group temperature difference was less than 0.18 K. During the slow cool down, the helium flows in the 5 K shield circuit and 45 K shield circuit were kept constant. The cool down rate was ~3 K/hour. After all cavities went through the superconducting transition, cool down was accelerated to allow efficient cryogenic operations.

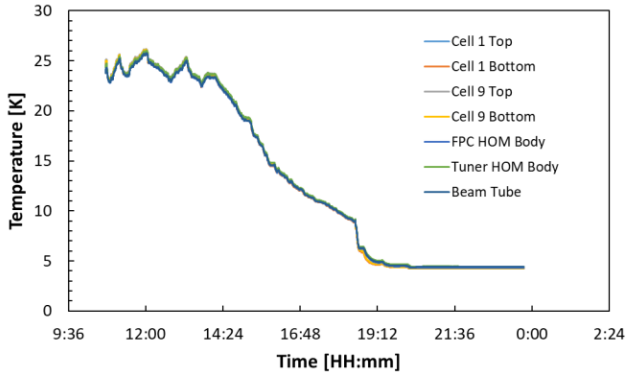

FIG. 5. Temperatures of cavity #1 during slow cool down.

### B. Magnetometer Readings During Cool Down Process

During the slow cool down, magnetic fields remained relatively stable except during the transition where the Meissner effect changed the local field slightly. The field spike due to thermal currents that was observed during fast cool down was absent during the slow cool down as shown on FIG 6. Table II lists the magnetometer readings right before cavity started superconducting transitions. It is important to note that the remaining field measured at the cavities during superconducting transitions were virtually the same as or smaller than the fields observed during the fast cool down. The significant spike in magnetic field due to fast cool down decreased to numbers comparable to those of slow cool down.

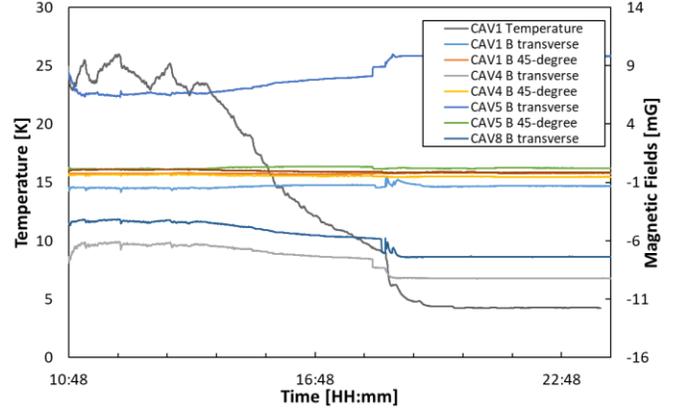

FIG. 6. Magnetometer readings during slow cool down.

TABLE II. Magnetic fields compared at the beginning of the superconducting transition during fast and slow cool downs. $B_T$ is the field measured by the transverse sensor and $B_{45}$ is the field measured by the 45-degree sensor.

| Cavity number | $B_T$ [mG] fast cool | $B_{45}$ [mG] fast cool | $B_T$ [mG] slow cool | $B_{45}$ [mG] slow cool |
|---|---|---|---|---|
| 1 | 0.02 | 0.9 | 1.4 | 0.2 |
| 4 | 6.8 | 1.2 | 7.8 | 0.5 |
| 5 | 7.9 | 0.4 | 8.1 | 0.2 |
| 8 | 1.2 | 0.7 | 5.9 | 0.2 |

### IV. Cavity Gradient Measurement

Cavity accelerating gradients were measured using two different methods and compared to assess the error distribution. One method measures forward power $P_f$ and external quality factor $Q_{ext}$ of the power coupler. The gradient can be calculated using equation

$$E_{acc} = \frac{1}{L}\sqrt{4P_f Q_{ext}\left(\frac{R}{Q}\right)} \quad (1)$$

where R/Q is the geometric shunt impedance of the cavity which is calculated to be 1012 Ω and L is the cavity effective length 1.038 meters.

$Q_{ext}$ is determined using time decay constant of the transmitted power. The forward power from RF amplifier to a cavity is calibrated through calorimetric load at the interface of cryomodule's input coupler. The power loss from coupler external joint to the cavity coupler flange port was calculated to be less than one watt and thus ignored. The forward power is measured at the output of a solid-state amplifier where the reflected power is minimized using an RF circulator placed before the power measurement location. This approach avoids the complication caused by imperfect directivity of an RF directional coupler under full reflection from a strongly overcoupled superconducting cavity. During continuous wave measurement, the power coupler antenna heats up slowly, and the external Q varies by 5% to 10% as the coupler inner conductor thermally expands and moves further toward the cavity beam axis. The forward power measurement is usually recorded together with the external Q.

Alternatively, the cavity gradient can be calculated using the transmitted RF power from a field probe $P_t$ and the external quality factor $Q_t$ of the field probe.

$$E_{acc} = \frac{1}{L}\sqrt{P_t Q_t \left(\frac{R}{Q}\right)} \quad (2)$$

In this case, the external Q of the field probe is measured during the vertical qualification test of each cavity, when a relatively well-matched power coupler is used to measure the field probe external Q. It is reasonable to assume that the external Q of field probe does not change since each cavity was directly transported to string assembly facility without the field probe being replaced. Care was taken to avoid any potential stress exerted onto the field probe during the cryomodule assembly. It was demonstrated through repeated vertical tests that the thermal excursions between room temperature and 2 K do not affect the probe's external Q at 2 K.

For the prototype cryomodule, gradients were measured for all eight cavities using the two methods described above. The gradients agreed within 0.5%.

## V. Cavity Q Measurement

The cavity intrinsic Q is calculated from the measured heat load $H$ and the accelerating gradient $E_{acc}$ using equation 3:

$$Q_0 = \frac{(E_{acc}L)^2}{\frac{R}{Q}H} \quad (3)$$

The heat load is measured calorimetrically via helium gas mass flow. A mass flow meter is installed at the discharge side of the helium pumping line that captures liquid helium boil-off. The flow meter is calibrated using a heater built into the cryomodule's liquid helium system. The calibration is performed at several heater power levels to obtain a coefficient of proportionality between the heater power and mass flow meter signal. The coefficient was found to remain approximately constant to allow linear interpolation of the amount of equivalent heat load. With the two-phase supply pressure, supply temperature, and supply valve fixed, the coefficient is expected to remain the same. To minimize potential non-linearity due to a complex cryogenic environment, the heater power during the flow meter calibration was adjusted to match the cavity heat load as closely as possible using mass flow as an indicator.

After the coefficient between the heater power and the change of mass was obtained, the cavity was powered in CW mode while the heater was turned off. The change of mass flow was then used to calculate the power loss of the cavity. FIG 7 illustrates a cavity heat load measurement at 16 MV/m.

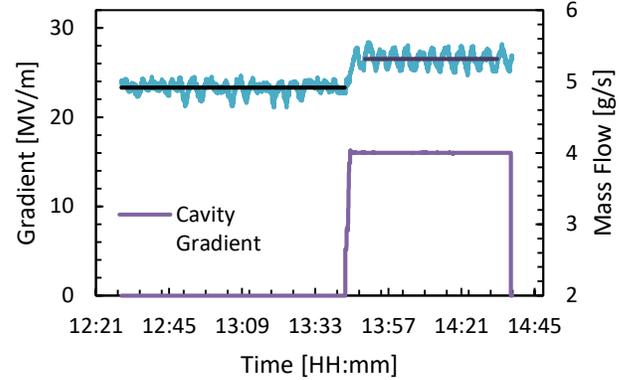

FIG. 7. Cavity dynamic mass flow measured for a gradient of 16 MV/m.

It should be noted that during the cryomodule test, microphonics detuning was measured that would be expected to be higher than could be compensated with the LCLS-II RF power supplies for a fixed drive frequency. This did not affect the measurement as the cavities were operated with a self-excited loop, but design modifications were later implemented to bring the microphonics detuning down to a level acceptable for future operation in the linac. The results and design modifications will be addressed in a future article.

## VI. Cavity Measurement Results

Table III lists the $Q_0$ and maximum accelerating gradient measurement results for eight cavities in the pre-production cryomodule compared to their $Q_0$ measured during vertical tests.

TABLE III. Cavity $Q_0$ measured compared to vertical test results

| Cavity number | Maximum Gradient | $Q_0$ measured in cryomodule at 16 MV/m* | $Q_0$ measured during vertical test at 16 MV/m |
|---|---|---|---|
| 1 | 21.2±1.1 | (2.6±0.3)x10$^{10}$ | (3.1±0.3)x10$^{10}$ |
| 2 | 19.0±1.0 | (3.1±0.3)x10$^{10}$ | (2.8±0.4)x10$^{10}$ |
| 3 | 19.8±1.0 | (3.6±0.4)x10$^{10}$ | (2.6±0.4)x10$^{10}$ |
| 4 | 21.0±1.1 | (3.1±0.3)x10$^{10}$ | (3.0±0.4)x10$^{10}$ |
| 5 | 14.9±0.7 | (2.6±0.2)x10$^{10}$ | (2.8±0.4)x10$^{10}$ |
| 6 | 17.1±0.8 | (3.3±0.4)x10$^{10}$ | (2.8±0.4)x10$^{10}$ |
| 7 | 20.0±1.0 | (3.3±0.3)x10$^{10}$ | (2.8±0.5)x10$^{10}$ |
| 8 | 20.0±1.0 | (2.2±0.2)x10$^{10}$ | (2.8±0.3)x10$^{10}$ |
| Average | 19.1±1.0 | (3.0±0.3)x10$^{10}$ | (2.8±0.4)x10$^{10}$ |

*Note: CAV5 $Q_0$ was measured at 14 MV/m.

The gradient of cavity #5 was limited by field emission and hence the $Q_0$ data was taken at 14 MV/m. The cavity #2 and cavity #6 gradients were limited by quench. Gradients of the other five cavities were limited by administrative restrictions.

The $Q_0$ result for each individual cavity showed some variations with respect to $Q_0$ in vertical test. Some of the eight cavities did have a stainless-steel flange which contributed to around 0.8 n$\Omega$ which may explain those cavity's variations compared to vertical test results. Nevertheless, the average cryomodule $Q_0$ agrees well with the vertical tested result within the measurement uncertainties.

## VII. Discussion

### A. Cryomodule Thermal Design and Thermo-Electric Current

One of the design goals for the cryomodule was to reduce the 2 K static heat load as much as possible. Two separate thermal boundaries were modified to minimize the heat load to 2 K helium circuit. One is a 5 K circuit used to intercept heat between the RF input coupler 2 K flanges connected to cavities and the higher temperature part of the coupler. It is also used to actively cool stepper motors since the motors heat up during tuner activation. Another thermal boundary is a 45 K thermal shield that intercepts the thermal radiation from the vacuum vessel at room temperature. The 45 K circuit is also used for heat intercepts on the input couplers. The couplers are the main source of heat leak as they connect the room temperature environment to the 2 K cavities.

Those thermal boundaries may not be consistent from cavity to cavity. During cryomodule assembly, thermal straps are attached from each circuit to cavities and couplers. Some variations of thermal conduction in those thermal straps are expected. The temperature difference between cavities are considered a possible source of elevated magnetic field before the superconducting transition as shown in FIG 4. We call this elevated magnetic field "static thermo-electric magnetic field."

During the fast cool down, additional thermo-electric currents can be induced by a high temperature difference between the dissimilar metal joints on either side of the cavity as shown in FIG 4. We name this elevated magnetic field "dynamic thermo-electric magnetic field." However, the temperature difference across the cavity becomes smaller as the cavity approaches to superconducting transition. While there is a sufficient temperature difference to expel the static thermo-electric magnetic field, the dynamic thermo-electric magnetic field becomes negligible. Other instrumented cavities experienced magnetic field similar to that in FIG 2. We conclude that the dynamic thermo-electric magnetic field is not harmful in our cool down procedure. The static thermo-electric magnetic field is considered to be present regardless of cool down rate, as it is caused by cryomodule's intrinsic temperature differences. This harmful field can be expelled by the temperature difference on the cavity if the temperature difference is sufficiently large and the material sufficiently strong magnetic flux expulsion behavior.

### B. Field Trapping and Field Expulsion

The thermo-electric magnetic field is expected to be fully trapped during the slow cool down, when there is negligible thermal gradient on the niobium cavity. $Q_0$ was measured for each cavity after 36 hours of soaking at 2 K. FIG 8 shows a general trend of improving $Q_0$ as the cool down mass flow increases.

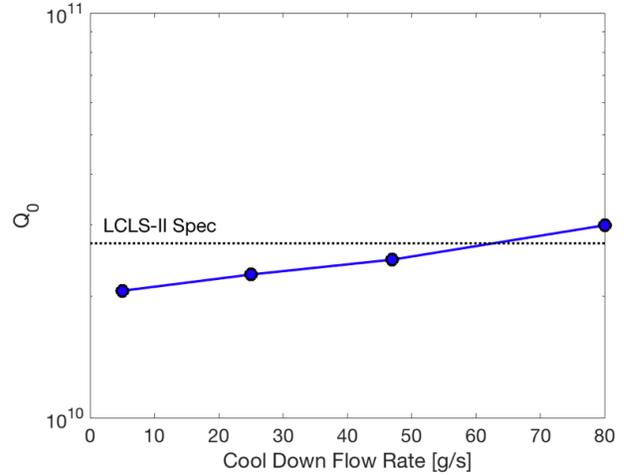

FIG 8: Q0 improves as the cool down mass flow increases.

A trapped field can be calculated under the assumption that all $Q_0$ degradation compared to vertical test is due to the trapped magnetic flux (after taking into account ~0.8 n$\Omega$ of surface resistance due to the stainless steel flanges), and using a coefficient between added surface resistance and trapped flux of 1.4 n$\Omega$/mG[13].

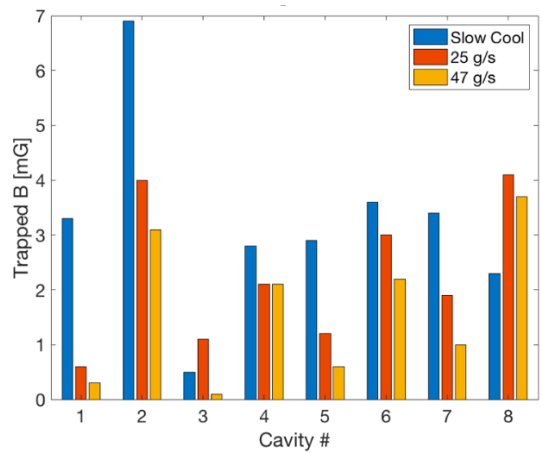

FIG 9: The trapped magnetic field calculated using cavity $Q_0$ and field sensitivity coefficient of 1.4 n$\Omega$/mG.

FIG 9 shows the trapped field calculated in this way. During the slow cool down, the trapped field is around 4 mG, and it decreases as the cool down mass flow increases, indicating improved magnetic flux expulsion due to the higher thermal gradient of the cavities.

## VIII. Summary

The Fermilab LCLS-II prototype cryomodule was tested, resulting in a record $Q_0$ exceeding the LCLS-II specification of $2.7 \cdot 10^{10}$. This measurement shows the significant impact of several new innovations implemented for the first time, including the use of the nitrogen doping cavity treatment and the use of high mass flow cooldown to expel magnetic flux. It also shows that the extremely high $Q_0$ of nitrogen doped cavities can be preserved from vertical tests to cryomodule assembly. Careful magnetic shielding design resulted in a historically low remnant field. The test result is an important design validation for the LCLS-II project and a significant milestone for developing SRF cryomodules operating in CW regime.

## IX. Acknowledgement

The completion of the first cryomodule for LCLS-II was an exemplary collaboration among four partner labs which are SLAC National Accelerator Laboratory, Thomas Jefferson National Accelerator Facility, Cornell University and Fermi National Accelerator Laboratory. The test team at Fermilab's Cryomodule Test Facility and cryogenic operational staff were exceptionally talented to support such a first of kind testing of a high performing CW cryomodule. S. Posen, A. Crawford and S. Belomestnykh provided editorial help and discussion.

This work was supported by the United States Department of Energy, Offices of Basic Energy Sciences and High Energy Physics. Fermilab is operated by Fermi Research Alliance, LLC under Contract No. DE-AC02-07CH11359 with the United States Department of Energy.